\documentclass[twocolumn,showpacs,preprintnumbers,amsmath,amssymb]{revtex4}

\usepackage{graphicx}
\usepackage{dcolumn}
\usepackage{bm}
\begin{document}

\preprint{APS/123-QED}

\title{Weak localization and antilocalization in semiconducting polymer sandwich devices}

\author{\"{O}. Mermer}
\affiliation{Department of Physics and Astronomy, The University of Iowa, Iowa City, IA 52242-1479}%
\author{G. Veeraraghavan, T. L. Francis}
\affiliation{Department of Electrical and Computer Engineering, The University of Iowa, Iowa City, IA 52242-1595}%
\author{M. Wohlgenannt}
\email{markus-wohlgenannt@uiowa.edu}
\affiliation{Department of Physics and Astronomy, The University of Iowa, Iowa City, IA 52242-1479}%

\date{\today}

\begin{abstract}
We have performed magnetoresistance measurements on polyfluorene sandwich devices in weak magnetic fields as a function of applied voltage, device temperature (10K to 300K), film thickness and electrode materials. We observed either negative or positive magnetoresistance, dependent mostly on the applied voltage, with a typical magnitude of several percent. The shape of the magnetoresistance curve is characteristic of weak localization and antilocalization. Using weak localization theory, we find that the phase-breaking length is relatively large even at room temperature, and spin-orbit interaction is a function of the applied electric field.
\end{abstract}

\maketitle

\section{Introduction}

The combination of quantum coherence and spin rotation can produce a number of interesting transport properties. In addition, numerous proposals for potentially revolutionary devices that use spin and spin-orbit (SO) coupling have appeared in recent years, including gate-controlled spin rotators \cite{SpinRotator} as well as sources and detectors of spin-polarized currents \cite{Spins}. Organic conjugated materials have been used to manufacture promising devices such as organic light-emitting diodes (OLEDs) \cite{ElectroluminescenceReview}, photovoltaic cells \cite{PVCellReview} and field-effect transistors \cite{reviewFET}. Recently there has been growing interest in spin \cite{nature,Baldo} and coherent quantum mechanical effects in these materials in order to assess the possibility of using them in spintronics and quantum information applications.

It is usually assumed that charge transport in polymeric semiconductor films occurs in the form of hopping transport of polarons and that the quantum mechanical phase varies randomly between sites. This is true for the strongly localized regime whose boundary is determined by the Ioffe-Regel criterion. In disordered inorganic conductors, however one often encounters weak localization (WL): coherent backscattering and the resulting interference of the quantum mechanical wavefunction of time-reversed trajectories leads to a conductance minimum. Strictly speaking this is only true in the spin-invariant case and a conductance maximum (weak antilocalization, WAL) occurs in the case of strong SO coupling \cite{HLN,Altshuler}. It is well-known from the study of diffusive transport in (inorganic) metals \cite{WeakLocalizationMetals}, semiconductor devices \cite{WeakLocalizationFET} and quantum dots \cite{WLQuantumDot} that phase-breaking and SO interaction times can be extracted from the WL and WAL traces, respectively, in magnetoresistance (MR) experiments. In fact MR and WL experiments have also been used to study organic materials such as carbon nanotubes \cite{WLNanotube}, nanotubules \cite{WLTubule} as well as heavily doped polymers \cite{WLPolymer}.

MR studies in heavily doped polymers found that WL can occur in stretch oriented high quality samples, but the measurements had to be performed at low temperature (below 4K) and in high magnetic fields (several Tesla). We have performed MR experiments in sandwich devices comprised of an undoped polymeric semiconductor, namely polyfluorene (PFO). We obtain several remarkable and unexpected results: we observe weak-field MR effects with a shape characteristic for WL and WAL effects \emph{even at room temperature}. In inorganic semiconductors such effects occur only at low temperatures (typically 4K and below). The measured WL cones are relatively narrow (in the mT range) indicative of relatively long phase-breaking times. At high applied electric fields, the WL cones change sign indicative of WAL and the importance of SO coupling.

Pi-conjugated polymer sandwich devices have been used extensively for OLEDs, photovoltaic cells and in transport studies, and we expect that the first organic spintronic devices will be of this structure. We have therefore studied WL and WAL in sandwich devices.

\section{Experimental}

Our thin film devices consist of the polymer PFO (poly(9,9-dioctylfluorenyl-2,7-diyl) end capped with N,N-Bis(4-methylphenyl)-4-aniline, see Fig. 1 inset) sandwiched between a top and bottom electrode. The polymer was purchased from American Dye Source, Inc. PFO was selected because of its common use in high quality OLEDs. The polymeric film was fabricated by spin-coating from toluene solution at 2000 rpm and baking at 90C overnight. For varying the film thickness, different concentrations were used, namely 7 to 30 mg/ml. The bottom electrode consisted of either ITO (indium-tin-oxide) covered glass or Au evaporated onto a glass slide. The top contact, either Al or Ca (covered by a capping layer of Al), was evaporated through a shadow mask at a base pressure of $10^{-6}$ mbar. The reported effects were observed in devices made from 2 different batches of PFO polymer. All manufacturing steps were performed inside a nitrogen glove-box.

The MR measurements were performed with the sample mounted on the cold finger of a closed-cycle He cryostat located between the poles of an electromagnet (GMW 3470). The temperature range between 10 K and 300 K was studied and the measurements were performed spanning the magnetic field range between $-100 mT < B < 100mT$, measured using a digital teslameter (3-DTM-133). The MR was determined by measuring the current at a constant voltage as a function of the applied field.  The current-voltage characteristics were measured using a Keithley 6517A.

\section{Experimental results and discussion}

\begin{figure}
\includegraphics[width=\columnwidth]{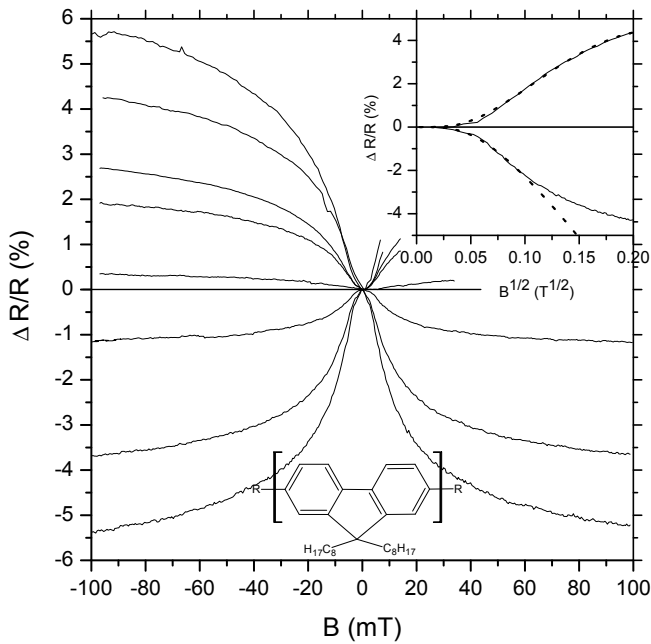}
\caption{\label{fig:Fig1} Typical examples of magnetoresistance, $\Delta R/R$ curves in PFO (see inset) sandwich devices. The figure is a collection of data measured at 200 K in a variety of different devices at different voltages (see text for discussion). The inset shows two examples for MR data plotted versus $B^{1/2}$, the dotted lines are fits to (equ.~\ref{eq:WL}).}
\end{figure}

Fig.~\ref{fig:Fig1} shows typical examples of the measured MR traces in our PFO sandwich devices. The figure is a collection of data measured at 200 K in a variety of different devices at different voltages. It is shown in Fig.~\ref{fig:Fig1} that both positive and negative MR is observed, dependent on operation conditions to be discussed later. \emph{Importantly, the observed MR traces closely resemble MR traces due to weak localization (WL, negative MR) and weak antilocalization (WAL, positive MR) well known from the study of diffusive transport in metals and semiconductors} \cite{WeakLocalizationMetals,WeakLocalizationFET,WLQuantumDot}. This suggests analyzing the MR data using the theory of weak localization. In the following we present strong evidence in support of this interpretation. We note that in principle there are several mechanisms yielding positive MR, such as Lorentz force, hopping magnetoresistance \cite{EfrosBook} and electron-electron interaction \cite{MRBook}, but we are unaware of another mechanism, other than WL, that gives weak-field negative MR.

\begin{figure}
\includegraphics[width=\columnwidth]{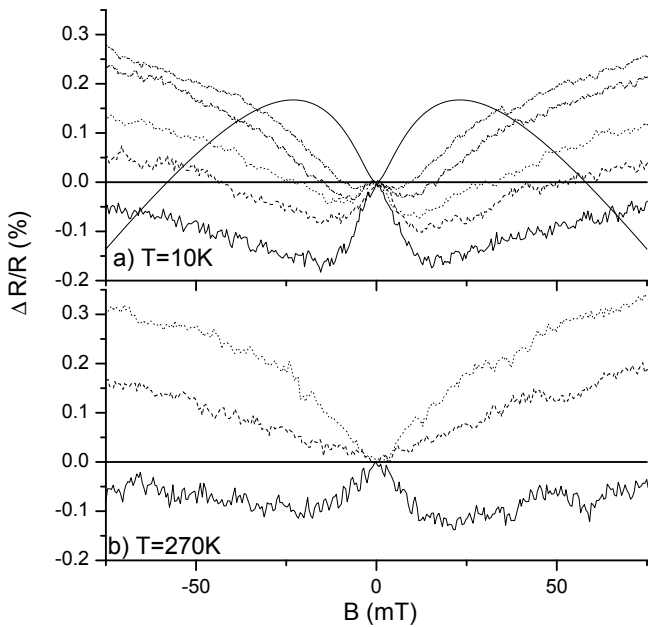}
\caption{\label{fig:Fig2} Examples of MR, $\Delta R/R$ traces close to the transition between WL and WAL in a Au/PFO ($\approx$ 150 nm)/Al device measured at 10 K (panel a)) and 270 K (panel b)). The solid line is calculated using (equ.~\ref{eq:WL}).}
\end{figure}

The MR due to quantum interference depends on three characteristic field values

\begin{equation}
B_{tr}=\frac{\hbar}{4eD\tau}, B_{\varphi}=\frac{\hbar}{4eD\tau_\varphi}, B_{SO}=\frac{\hbar}{4eD\tau_{SO}} \label{eq:widthofHLN}
\end{equation}

where D is the diffusion constant, $\tau$, $\tau_\varphi$ and $\tau_{SO}$ are the elastic scattering time, the phase-breaking time and the SO relaxation time, respectively. Additional spin-depending scattering mechanisms may also contribute, but they will be omitted here for simplicity. Quantum interference occurs when $B_{tr} \gg B_\varphi$. The MR will be negative (WL) when $B_\varphi \gg B_{SO}$, and positive (WAL) in the opposite case. In the regime $B_\varphi \approx B_{SO}$ the MR can change sign as function of B. Fig.~\ref{fig:Fig2} shows the measured MR traces in the transition region between WL and WAL. We note that very similar transition region traces are also found in (inorganic) metals and semiconductors \cite{WeakLocalizationMetals,WeakLocalizationFET,WLQuantumDot}.

The exact formula describing the MR in the quantum interference regime depends on the dimensionality of the system. We find that we are dealing with three-dimensional quantum interference. In two-dimensional systems only in-plane looped paths are possible, and therefore only a transverse magnetic field results in enclosed magnetic flux and hence phase difference. We find, however that the MR traces are independent of the angle between film plane and applied field. All measurements shown in the figures were performed with an in-plane magnetic field.

According to Al'tshuler et al. \cite{Altshuler} the MR of a three-dimensional metallic system with SO interaction is given by:

\begin{eqnarray}
& \frac{\Delta R}{R^2} \propto \frac{\Delta \rho}{\rho^2}=\frac{e^2}{2\pi^2\hbar}\left (\frac{eB}{\hbar} \right )^{\frac{1}{2}} \left (\frac{1}{2}f_3\left (\frac{B}{B_\varphi} \right )-\frac{3}{2}f_3\left (\frac{B}{B_2} \right ) \right) \label{eq:WL}\\
& B_2=B_\varphi+\frac{4}{3} B_{SO}
\end{eqnarray}

The function $f_3$ is given in Ref. \cite{Altshuler}. In fitting our experimental data we have instead used an approximate expression for $f_3$ given by Baxter et al. \cite{Baxter}, which has been shown to be accurate to better than 0.1\% for all arguments.

\begin{eqnarray}
&& f_3(z) \approx 2 \left [\left (2+\frac{1}{z} \right )^{ \frac{1}{2}}- \left (\frac{1}{z} \right )^{\frac{1}{2}} \right ] -\\
&& - \left [\left (\frac{1}{2}+\frac{1}{z} \right )^{-\frac{1}{2}}+\left (\frac{3}{2}+\frac{1}{z} \right )^{-\frac{1}{2}} \right ] + \frac{1}{48} \left (2.03+\frac{1}{z} \right )^{-\frac{3}{2}} \nonumber
\end{eqnarray}

We succeeded to fit both the WL and WAL traces shown in Fig.~\ref{fig:Fig1} with (equ.~\ref{eq:WL}) using a common $B_\varphi=1 mT$, but with different values for $B_{SO}$. $B_\varphi=1 mT$ translates into a phase-breaking length, $l_\varphi = \sqrt{D\tau_\varphi} \approx 400 nm$. Examples of the resulting fits are shown as the dotted curves in Fig.~\ref{fig:Fig1}, inset yielding $B_{SO} = 0$ and 15 mT for WL and WAL, respectively. We chose to fit the MR data plotted as a function of $B^{1/2}$ as it is commonly done in the literature because it results in a more reliable fitting procedure. The quality of the fits are satisfactory although significant deviations between fit and experimental data occur in the high field region. We note, however, that similar deviations are commonly found in WL and WAL studies in the literature \cite{WeakLocalizationFET,WLQuantumDot}. We may therefore conclude that \emph{both} WL and WAL traces can be understood using weak localization theory.

\begin{figure}
\includegraphics[width=\columnwidth]{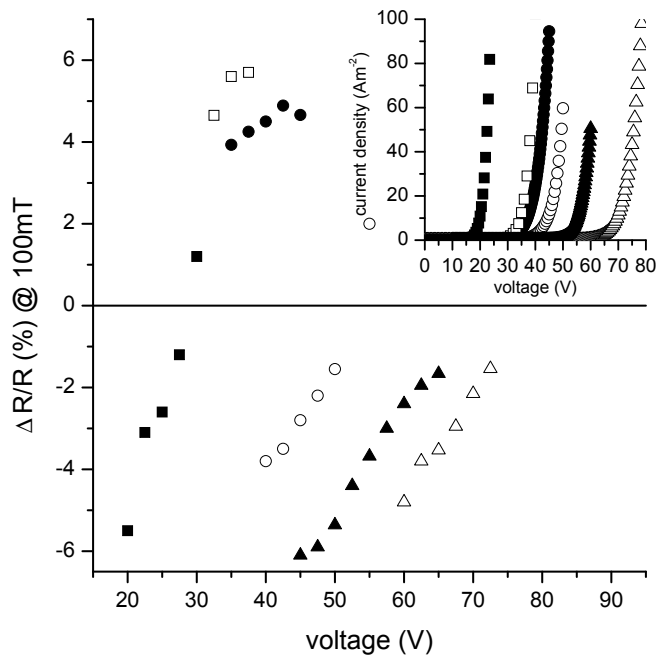}
\caption{\label{fig:Fig3} Dependence of the magnitude of the MR effect, $\Delta R/R$ at 100 mT and 200 K on the device voltage in a variety of devices with different polymer film thickness and electrode materials. The inset shows the current-voltage characteristics of these devices. $\blacksquare$ is for an ITO/PFO ($\approx$ 60 nm)/Ca device, $\square$ is for ITO/PFO ($\approx$ 100 nm)/Al, $\bullet$ is for ITO/PFO ($\approx$ 100 nm)/Ca, $\circ$ is for ITO/PFO ($\approx$ 140 nm)/Ca, $\blacktriangle$ is for Au/PFO ($\approx$ 150 nm)/Ca, and $\vartriangle$ is for ITO/PFO ($\approx$ 300 nm)/Ca.}
\end{figure}

We generally found the observed MR effect to be much smaller in magnitude than that predicted by (equ.~\ref{eq:WL}). However, this discrepancy is not surprising since our devices show strongly non-linear IV characteristics (see Fig.~\ref{fig:Fig3}, inset) - as is generally the case in organic sandwich devices - whereas (equ.~\ref{eq:WL}) is derived for ohmic resistors. In addition, $\Delta R/R \propto R$ should hold true according to (equ.~\ref{eq:WL}), but we find that the resistance of our devices decreases much faster with increasing voltage than does the magnitude of the MR effect. Since $\Delta R/R \propto R$, negative MR effects larger than 100\% would be predicted for the very large R's of our devices (typically $100 k\Omega$). The observed saturation in $\Delta R/R$ must obviously occur. The most striking discrepancy between the prediction of (equ.~\ref{eq:WL}) and our experimental data occurs in the transition region data shown in Fig.~\ref{fig:Fig2}. (Equ.~\ref{eq:WL}) predicts that the MR trace makes the transition from WL to WAL first at small fields (solid line in Fig.~\ref{fig:Fig2} a) with $B_\varphi = 1 mT$ and $B_{SO} = 7 mT$) whereas the experimental curve makes the transition first at large fields. We presently do not understand the reason for this discrepancy.

Fig.~\ref{fig:Fig3} shows the dependence of the magnitude of the MR effect at 100 mT and 200 K on the device voltage in a variety of devices with different polymer film thickness and electrode materials (details are given in the caption of Fig.~\ref{fig:Fig3}). ITO and Ca are commonly used in OLEDs since they result in relatively small barriers for hole and electron injection, respectively. Au often gives ohmic contacts for hole injection and Al is another commonly used top-electrode material. The current-voltage (IV) characteristics of the measured devices are shown as an inset to Fig.~\ref{fig:Fig3} in a linear-linear plot. We found that the linear-linear IV plot is determined mostly by the PFO film thickness, and that IV and MR curves do not critically depend on the electrode materials used. Fig.~\ref{fig:Fig3} is in essence a plot of the magnitude of the backscattering cone as a function of the device voltage, and therefore - knowing the devices' IV characteristics - the device resistance. It is seen that $\Delta R/R$ increases in magnitude with increasing device resistance - i.e. decreasing voltage - in qualitative agreement with (equ.~\ref{eq:WL}). The most striking result shown in Fig.~\ref{fig:Fig3} is that the transition between WL and WAL is apparently driven by the applied electric field. We note that the applied electric fields are very large in polymer sandwich devices (typically $10^{6} V/cm$). In agreement with our finding, electric fields are known to cause the WAL effect \cite{SOElectricField,WALElectricField1,WALElectricField2,WALElectricField3,WALElectricField4}. A similar cross-over from WL to WAL has been observed in inorganic metal thin films when covering the film with increasing amounts of Au, known to be an efficient SO coupler \cite{WeakLocalizationMetals} and as a function of gate voltage in inorganic devices \cite{WLQuantumDot}.

\begin{figure}
\includegraphics[width=\columnwidth]{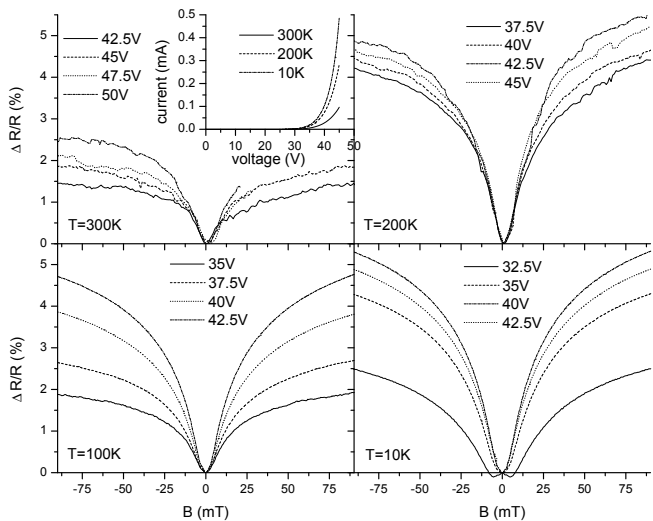}
\caption{\label{fig:Fig4} Examples of typical "antilocalization" magnetoresistance, $\Delta R/R$ curves in an ITO/PFO ($\approx 150 nm$)/Ca/Al device measured at different temperatures, namely 10 K, 100 K, 200 K, and 300 K. The applied voltages are assigned. The inset shows the IV characteristics at different temperatures.}
\end{figure}

Fig.~\ref{fig:Fig4} shows measured WAL cones as a function of temperature, similar results were obtained for WL traces. It is seen that the MR effect is clearly observed at all temperatures spanning the range between 10K and 300K. We find that the magnitude and width of the backscattering cones are relatively insensitive to temperature. This is surprising, since WL has previously only been observed at low temperatures, and because D and $\tau_\varphi$ are expected to be strong functions of temperature. However, D is expected to increase with temperature whereas $\tau_\varphi$ is expected to decrease, their temperature dependencies may therefore partially cancel. In addition we note that the device's IV characteristics (see Fig.~\ref{fig:Fig4}, inset) is also only weakly temperature dependent. We find that the device resistance even slightly decreases with decreasing temperature, in stark contrast to the expectation for hopping transport. We take this as a sign of the devices' high quality. We find that $B_\varphi$ in Fig.~\ref{fig:Fig4} increases from about 1 mT ($l_\varphi \approx 400 nm$) at 10K to about 5 mT ($l_\varphi \approx 150 nm$) at 300K. The decrease in magnitude of $\Delta R/R$ between 200K and 300K also indicates a decline in $l_\varphi$ between these two temperatures. Finally we note that since D is known to be relatively small in pi-conjugated polymers, $\tau_\varphi$ may be very long in these materials, possibly in the ns range; and that backscattering effects may be enhanced in semiconducting polymer films  because polymer chains are often coiled up.

In summary, we found strong evidence for the occurrence of quantum interference corrections to the charge carrier transport in polymeric sandwich devices. To the best of our knowledge, this is the first observation of this kind. Quantum interference corrections occur only at low temperatures (typically 1 K or even below) in (inorganic) metals \cite{WeakLocalizationMetals}, high-mobility semiconductor heterojunction devices \cite{WeakLocalizationFET} and quantum dot devices \cite{WLQuantumDot}. In stark contrast we found WL traces even at room temperature, with virtually undiminished magnitude and width. Our discovery indicates that the quantum mechanical phase may remain highly correlated over many hops in high-quality polymer sandwich devices, even at room-temperature. In addition we found that the SO interaction strength is a function of the electric field and this finding may have important implications to the currently developing field of organic spintronics.

\bibliography{weaklocalization}

\end{document}